\documentclass[twocolumn]{openjournal}

\usepackage[T1]{fontenc}
\usepackage{ae,aecompl}
\usepackage{natbib}
\usepackage{wasysym}

\usepackage{xcolor}
\begin{document}
\title{Generating eccentricity from envelope stripping in the Radius Valley}
\author{Brad M. S. Hansen}
\email[email]{ hansen@astro.ucla.edu}
\affil{ Mani L. Bhaumik Institute for Theoretical Physics,  Department of Physics \& Astronomy, University of California Los Angeles, \\ Los Angeles, CA 90095}

\begin{abstract}
We demonstrate that the stripping of planetary envelopes can result in the excitation of planetary orbital eccentricity if the stripped gas remains in the system long
enough to interact gravitationally and absorb angular momentum from the planetary orbit. We estimate that envelope mass fractions of a few percent can
excite  eccentricities
$\sim 0.1$ for single planets. This process can potentially explain the elevated eccentricities observed for planets whose radii lie in, or near, the radius valley.

In multiple planet systems, the stripped gas can mediate angular momentum exchange between neighbouring planets, causing the separation to expand.
Using data from the Kepler satellite, we find that planet pairs, whose members straddle the radius valley, do appear to be systematically wider than those in
which both members are sub-Neptunes.

These results support the idea that planets in, or just below, the radius valley are the stripped cores of planets that have lost mass, because the elevated
eccentricities represent evidence that the gas remained in the system long enough to absorb additional angular momentum from the planet.

 \end{abstract}

\maketitle



\section{Introduction}

The exoplanet census performed by the Kepler space telescope \citep{Kep10} has revealed that the most common type of planet in the
Galaxy is a planet a few times the mass of Earth, with an orbital period of a few days to a few tens of days \citep{HMJ10,MM11,HMB12,DZ13,FTC13,MPA18,TCH18,ZCH19}. Such planets
are not present in our own Solar system and so we must infer their natures by remote observation alone. Observation
of planetary radii from transit observations, allied with mass constraints from either transit time variations or radial
velocities, suggest that there are two broad classes of rocky planets, with some appearing to be essentially bare, rocky, cores,
and some seeming to possess low density envelopes comprising a few percent of the total mass \citep{LF14,WM14,JHF16}, although
there remains an active debate as to the relative contributions of water or Hydrogen/Helium envelopes to the total mass
inventory \citep{KFS20,VGH20,ZJH21}.

The source of the lower-density envelopes of many of these planets is still uncertain, which can be affected by the accretion
of volatile-rich material that condensed beyond the snow line \citep{ZJS19,MDA20} or by accretion of Hydrogen gas from the nebula during the protoplanetary
disk phase \citep{Raf06,RBL11,IH12}. Furthermore, there is evidence that some planets can lose their low density envelopes over time,  
 either because of the photoevaporation of the low density envelope by
high energy radiation from the star \citep{OW13,JMP14,HB15,CR16}, or by the thermal evaporation powered by residual heat of accretion in the planetary
core\citep{GSS16,GS19}. This claim is supported by the observation of a `radius gap' \citep{FPH17,VAL18,PRI22}, indicating a bifurcation in the properties
of the modern planetary population, although this may be affected by the initial conditions of planet formation \citep{VGH20}.

A potentially important insight into the physical origin of this radius gap is the observation \citep{GPE25} that the orbits of planets which lie within
the range 1.75--1.93~$R_{\oplus}$ show an elevated level of eccentricity compared to those  of planets both below and above this 
radius range. It suggests that the process by which mass is removed from the planets is also responsible for dynamical excitation.
In this paper we discuss the hypothesis that the evaporation of gaseous envelopes from low-mass planets does not remove the
gas from the system immediately, and that torques acting between the planet and the stripped envelope remove angular momentum from
the planet, thereby exciting its eccentricity.
In \S~\ref{Torus} we describe how
gravitational interactions between the planet and the stripped gas can extract angular momentum from the planetary orbit and excite
eccentricity. We examine the degree to which tidal damping may affect the resulting eccentricity in \S~\ref{Tidal}. The
stripped gas may also mediate angular momentum exchange between neighbouring planets, which we discuss in \S~\ref{Multis}
and \S~\ref{Discuss} discusses the broader implications of this model.

\section{A stripped planetary envelope becomes a gaseous torus}
\label{Torus}

Planets that orbit close to their parent stars experience a much higher level of instellation than the Solar system planets.
The resulting energy input can heat the gas in the atmosphere to the point that it drives mass loss from the planet.
The resulting outflow has been observed from gaseous giant planets \citep{VLD03,LBW12} as well as Neptune-mass
planets \citep{EBW15,LEB17,LSO25}.  The observed rates are not sufficient to evaporate the extensive envelopes of gas
giants, but are enough to strip the Hydrogen envelopes from some super-Earths \citep{OW13}, and thus may be responsible
for the presence of the ``Radius Valley'' \citep{FPH17}.

In models of this planetary atmosphere stripping, the   loss of the Hydrogen--Helium envelope is driven by the heating of the planetary envelope,
either by the irradiation from the central
star, or by residual thermal heat of formation. The energy deposition heats the gas in the atmosphere to the point it overflows the Hill
sphere \citep{MCM09,OW13,JMP14,HB15,CR16}. However, for those planets to experience mass loss, 
the required escape velocity from the planetary Hill sphere is only
 $\sim \left(  M_p/M_*\right)^{1/3} $
that of the escape velocity from stellar potential well --  only a few percent for Neptune-like masses. 

In the absence of other physical effects, the gas lost from the planet will simply go into orbit around the host star. Indeed,
the star WASP-12b is argued to be entirely encircled by a gaseous torus stripped from the planet \citep{HFA12,FAH13}, and the
GJ436 system also shows absorption over a significant fraction of the orbit \citep{LEB17}.

Radiation pressure and ram pressure from the stellar wind can affect the stripped gas as well.
The precise nature and geometry of the planetary outflows is known to depend on the strength of planetary and stellar
magnetic fields  and on the hydrodynamic interaction between the planetary outflow and the stellar wind \citep{CKD11,MUK15,CNF17,MMK19,MO22}.
If the stellar wind is weak, the gas forms a long-lasting, dense torus, compressed by the ram pressure of the wind \citep{DCF18} and
some of the material may even accrete back onto the star. Stronger winds first remove the inward moving gas, and
collimate the outflow into a cometary tail configuration, which becomes more radial as the stellar wind increases \citep{MMK19}.

\subsection{Gravitational Interaction with the Torus}

Thus, in many -- possibly most -- of the mass-losing systems, the gas stripped from the planet remains present in
the system for many dynamical timescales. If this process produces a long-lived gaseous torus, then the torus will experience a varying gravitational perturbation from the potential of the orbiting
planet. The interaction between a planet and the natal gaseous protoplanetary disk \citep{GT80,Ward97} is known to launch waves in the disk, which transfer
energy and angular momentum between planet and disk (see \cite{PDD23} for a review).  A similar interaction is to be expected in this case, although the surface density distribution
of the gas is different, and the mass in gas is considerably smaller. Furthermore, moderate-to-strong stellar winds are expected to produce a torus or cometary tail that lies
primarily exterior to the planetary orbit. Thus, we model this by considering 
 the effect on the planetary orbit of a low mass gas torus, located exterior to the planetary orbit.

The reaction of a gas disk to this periodic forcing will be determined by the structure of the disk, and the strength, and saturation, of co-orbital, co-rotation and eccentric
Lindblad resonances \citep{GS03}. A torus exterior to the planet will not be affected by co-orbital resonances, and the eccentricity excitation of the planet will be
determined by the competition between external co-rotation and eccentric Lindblad resonances. 
In an extended protoplanary disk, as 
 discussed in \cite{GS03}, the launching of waves in a gas disk will damp the planetary eccentricities as long as both the Lindblad
and corotation resonances are fully excited, but that the saturation of co-rotation resonances weakens their influence and the waves launched
from the Lindblad resonances can lead to eccentricity excitation. The
saturation \citep{OL03} is controlled by a parameter
\begin{eqnarray}
p & = &  \left(\frac{r}{h} \right)^{2/9} \left( \frac{M_*}{M_p} \right)^{1/9} \frac{e_p}{\alpha^{1/4}} \\
&\sim& 2.465 \left( \frac{M_p}{10 M_{\oplus}} \right)^{-5/27}
\left( \frac{\alpha}{0.01} \right)^{-1/9}  \frac{e_p}{0.01}
\end{eqnarray}
where we have assumed that the aspect ratio of the torus ($h/r$) is set by the Hill sphere radius of the planet. The disk viscosity is represented by $\alpha$, and 
$e_p$ is the planetary eccentricity. The critical threshold for eccentricity excitation is $p>0.157$ (using the saturation formula
of \cite{GS03}). Thus, we see that we expect that interaction with the stripped material for these parameters should pump the
planetary eccentricities. This mechanism does require a small, but finite, initial  planetary eccentricity but 
transit timing observations suggest that residual eccentricities $\sim 0.01$ are common in these
systems \citep{HL17} -- sufficient to provide the initial excitation. Torques can induce both eccentricity excitation and orbital migration,  but the
timescale for eccentricity excitation is shorter \citep{GS03}, so we will assume the planet remains in place, but that the eccentricity is increased
as angular momentum is extracted. Similar evolution is seen in the case of stellar binaries interacting with circumstellar disks \citep{SWH}, although
the mass ratios are higher in that application.

To estimate the timescale for eccentricity excitation, we adapt the expression for the eccentricity pumping time \citep{GT80,GS03}, by approximating
the gas as a narrow ring, offset by a distance $w$ from the planetary semi-major axis $a_p$, so that 
\begin{equation}
\frac{1}{t_e} = \frac{d\ln e_p}{dt} = G(p) f \left( \frac{M_p}{M_*}  \right)^2 \left( 1 + \frac{a_p}{w} \right)^4  \Omega \label{te}
\end{equation}
where $G(p)$ is determined by the saturation formula (and $G(p)=1.3$ for $p=2.465$). The Keplerian
frequency at the planet semi-major axis is represented by $\Omega$, and $f$ is the fraction of the original planet mass that is contained in the gas ring.
For a planet of mass 10 $M_{\oplus}$ and a stripped mass fraction of $f=0.01$, this expression yields
$ t_e \sim 6000 P$, where $P$ is the planetary orbital period, and we have taken $w$ to be  comparable to the Hill sphere
radius. This amounts to a timescale $\sim 100$~years for a planet with an orbital period of 10~days. This is shorter
than most characteristic mass loss timescales ($t_0 \sim $1--100~Myr) and so the ring is unlikely to contain the full
mass of the envelope. If we approximate $f = 0.01 (t/t_0)$, then a more realistic characteristic timescale for eccentricity
growth is $t \sim 10^4$ years.

Thus, we expect rapid excitation of planetary eccentricity even with a small amount of mass retained in orbit
around the planet. If the envelopes that evaporate from planets in the radius valley are not removed immediately from
the neighbourhood of the orbit, we anticipate that planetary eccentricities should be excited for those planets moving
through the radius valley towards smaller radii.

\subsection{Long Term Outcome}

However, an important limitation of this growth is that the mass in the envelope is much smaller than that of a traditional
protoplanetary disk. The calculation of migration or eccentricity excitation in protoplanetary disks is usually calculated under
the assumption that the disk is an infinite sink of energy and angular momentum \citep{GR01,Raf02}. 
In that event, the dissipation of the torques dumps angular momentum into the gas, driving it outwards. Viscous evolution
of the external disk drives mass inwards, and the competition between these processes determines the depth, shape and
extent of the gap that opens around a planet.
In our case, there is no external reservoir of gas to replenish the mass that is driven out. Thus, we expect that the gaseous
ring will be driven outwards as it absorbs angular momentum from the planet and that the coupling between gas and planet
will weaken as they become more seperated.

To estimate the limit of this process, we note that that the expression in equation~(\ref{te}) falls off sharply as $w$ increases. Thus,
if we assume that the ring absorbs angular momentum and moves outwards, the coupling to the planet will weaken
rapidly. Furthermore, the eccentricity excitation is driven by waves launched at outer eccentric Lindblad resonances, located at offsets
$w = \left( \left[(m+1)/(m-1)\right]^{2/3} - 1 \right) a_p $ for integer order m.
For a gas torus whose width is comparable to the original Hill diameter, we expect the torus will have width $\sim 0.043 a_p (M/10  M_{\oplus})^{1/3}$.
The spacing between commensurabilities is smaller than this close to the planet, so that the torus always contains a Lindblad resonance
and will be driven outwards. However, farther from the planet, the spacing between commensurabilities increases, and becomes wider
than the torus for $m<6$. Depending on the degree 
of viscous spreading of the torus as it expands outwards, the coupling could extend to m=5 perhaps, but we expect
the coupling to weaken once the separation reaches this scale. This will stall 
,the outward evolution of the torus and the eccentricity excitation of the planet.
Therefore,  assuming that the eccentricity excitation stalls  when m=5, this yields
$w/a \sim 0.3$, $1 +a/w \sim 4.2$ and $t_e \sim 1.5 {\rm Myr}$ for a $10 M_{\oplus}$ planet in a 10 day orbit,
that lost a mass fraction $f=0.01$.

\begin{figure}
\centering
\includegraphics[width=1.0\linewidth]{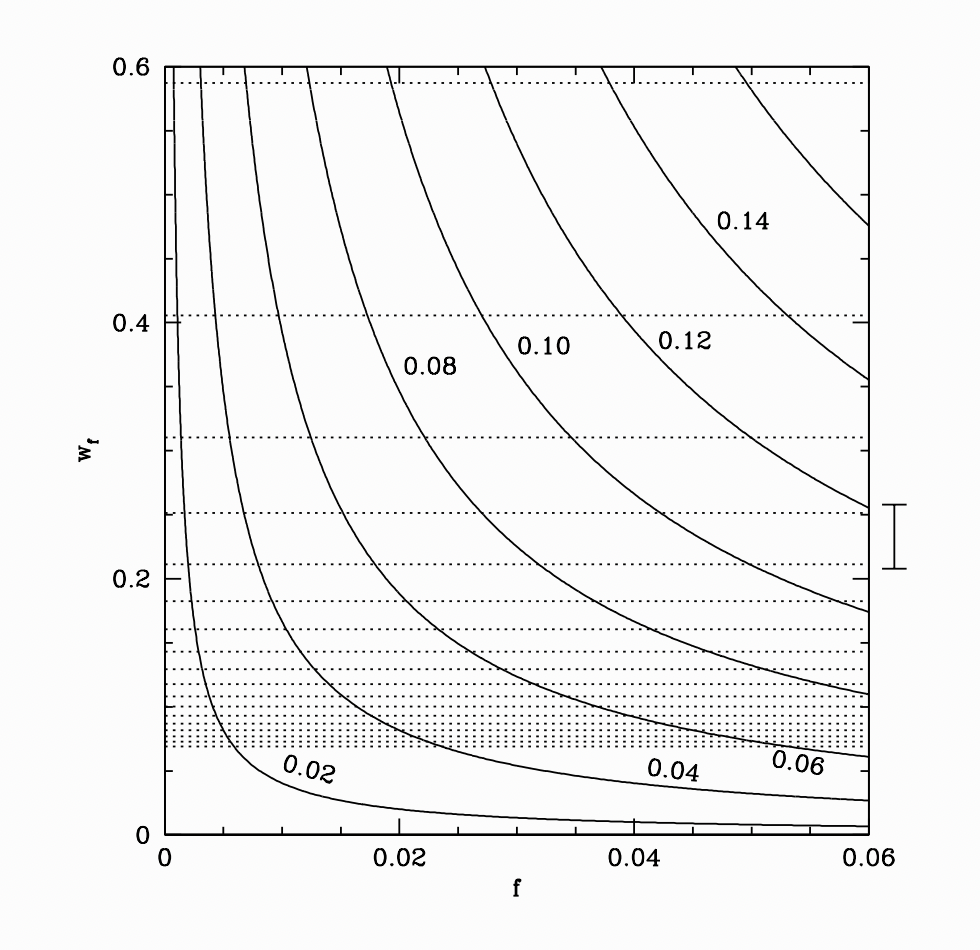}
\caption{The solid contours  indicate contours of constant final planetary eccentricity, given the amount of 
mass lost ($f$) and the separation to which the ring migrates ($w_f$). The horizontal dotted lines indicate
the locations of first order Outer Lindblad resonances
(the density of such resonances increases towards the bottom of the plot and the dense web at the bottom
has been omitted for clarity). The error bar on the right side of the figure indicates the extent of a torus
whose width is equivalent to the Hill sphere diameter of a Neptune-mass ($15 M_{\oplus}$) planet.
\label{outring}}
\end{figure}

To assess the limit of this process, let us assume 
  that the star loses an atmospheric mass fraction $f$, which absorbs angular momentum from the
  orbit and is driven outwards to an offset $w_f$ exterior to the planet location. To drive the ring of
  lost mass outwards requires the extraction of angular momentum from the planetary orbit, resulting
  in an eccentricity
  \begin{equation}
  e_p^2 = 2 f \left( (1 + w_f)^{1/2} - 1 \right) + \left[ 2 (1 + w_f)^{1/2} - 2 - w_f \right] f^2
  \end{equation}
  which reduces to $
  e_p \sim 0.056 \left( \frac{f}{0.01} \right)^{1/2} \left( \frac{w_f}{0.3} \right)^{1/2}$
  to first order in $f$ and $w_f$.
  Figure~\ref{outring} shows the more complete solution, demonstrating that one can achieve
  eccentricities $>0.1$ if the mass loss exceeds 5\% of the planetary mass.

\section{Tidal Circularisation}
\label{Tidal}

Many of the planets in the radius valley are close to the host star, and thus eccentricity excitation will be combatted
by eccentricity damping due to tides excited in the planet.
For a planet to retain a measureable eccentricity in a short period orbit, the tidal dissipation in the planet must
be sufficiently weak. The circularisation time for a low mass planet is \citep{JGB08}
\begin{equation}
\frac{\tau_e}{Q_p} = 2 \times 10^{7} {\rm yrs}  \left( \frac{M_p}{10 M_{\oplus}} \right)
\left( \frac{M_*}{M_{\odot}}\right)^{2/3} \left( \frac{R_{\oplus}}{R_p}\right)^{5}
\left( \frac{P}{\rm 5 \, days} \right)^{13/3} 
\end{equation}
where $Q_p$ is the tidal quality factor, which depends on the level of  dissipation within the planet,
and thus on the interior structure. Estimates for the tidal dissipation in Neptune \citep{BM92} suggest
$Q_p \sim 10^4$, which we take to represent a plausible value for the sub-Neptunes. Estimates
of tidal dissipation in the Earth are somewhat lower, with $Q_p \sim 10$ \citep{GS66,WSY78,CR91}. However, much
of this dissipation is believed to result from pelagic turbulence in the Earth's oceans \citep{Bell75,ER00},
which may not be present if the evaporated surfaces are dry. The estimate for dissipation in the solid
Earth alone, and for  a truly rocky planet
like Mars,  is $Q_p \sim 10^3$ \citep{REL01,LDP07}, which is what we will use.

\begin{figure}
\centering
\includegraphics[width=1.0\linewidth]{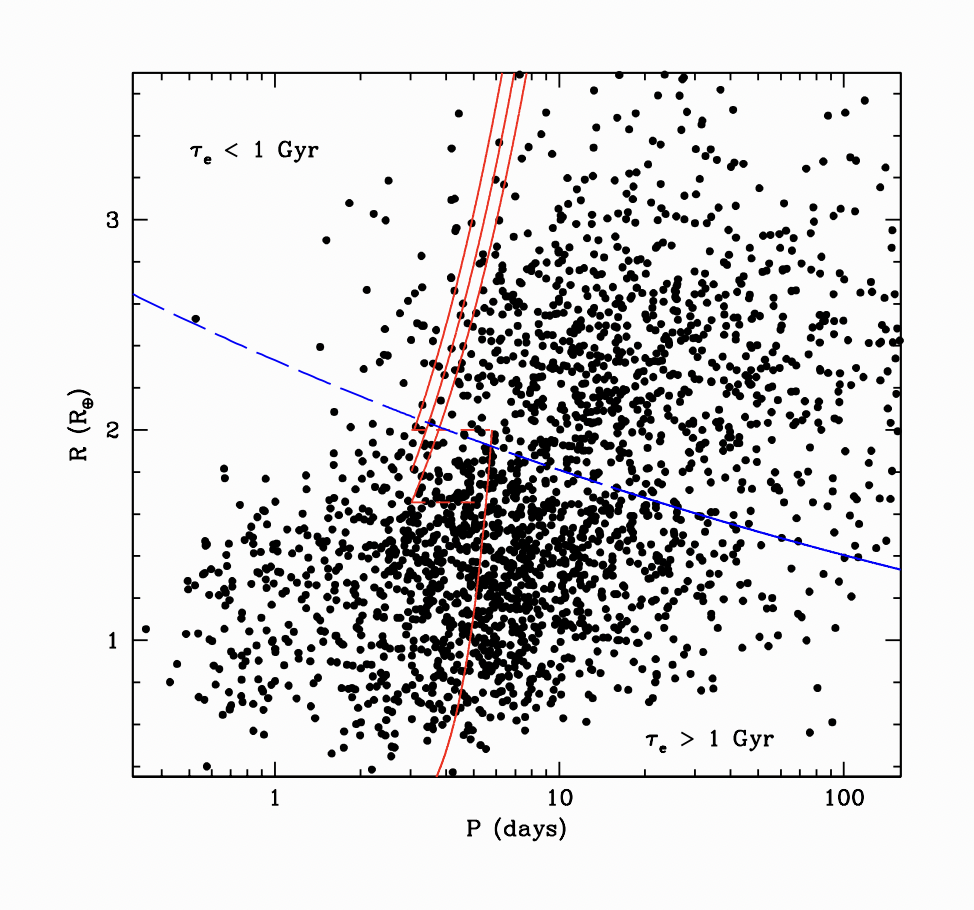}
\caption{The solid points represent planets from the catalog of  \cite{BHG20}. The
solid curves represent the criterion that $\tau_e = 10^9$~years, for two different values
of Q=$10^3$ (rocky planets) and Q=$10^4$ (Neptunes).  The transition from one limit to the
other, as the envelope evaporates, is uncertain. Therefore, the region between the two estimates 
represents a region of uncertainty -- planets that should be circularised if rocky, but not if they
retain a Hydrogen envelope.
 Planets to the left of the red curves are expected to have been
circularised, while those to the right may  be expected to retain some of their orbital eccentricity.
The blue dashed line represents the fit to the radius valley location \citep{PRI22,HV23}
\label{Tides}}
\end{figure}

Thus, in Figure~\ref{Tides}, we show the period--radius diagram for the planets in the 
Kepler catalogue, calibrated using GAIA stellar magnitudes \citep{BHG20}.
The observed values are  compared to two tidal circularisation thresholds, for rocky planets and sub-Neptunes,
shown as red curves. The rocky planet model is the silicate planet model taken from \cite{ZJH21}, assuming
a dry Earth $Q=10^3$, and extending from 0--10$M_{\oplus}$. We show three sub-Neptune models
from \cite{LF14} for an age of 1~Gyr and subject to ten times the terrestrial irradiation. These represent
three rocky core masses (5.5, 8.5 and 13 $M_{\odot}$), each with a range of Hydrogen envelope
masses. For these planets, we assume $Q=10^4$.

The distribution of planets in Figure~\ref{Tides} shows that the bulk of the observed population
have tidal circularisation times $\tau_e>$1Gyr. As such, it is plausible to observe finite eccentricities
in the radius gap -- we do not expect most of these planets to circularise. Furthermore, we see that
most of the $R>2 R_{\oplus}$ planets, and a substantial fraction of the $R<1.8 R_{\oplus}$ planets,
also lie to the right of the red curves. Thus, the absence of significant eccentricities in these two
populations is also important -- it indicates that the dynamical excitation of the planetary eccentricities
is directly related to their presence in the radius gap. If eccentricities were excited for planets above
and below the radius gap, they would be expected to mostly persist. We will return to this question in \S~\ref{Discuss}.

\section{Planetary Pairs straddling the Radius Gap}
\label{Multis}

The above calculations assume a single planet losing mass, and driving the stripped
mass to larger radii to absorb the angular momentum, with the process stalling once
the torus is driven to a large enough distance.
However, many of the planets
shown in Figure~\ref{Tides} are actually part of multiple planet systems. Therefore,
we must consider the effects of the above model when there is a second planet that
lies exterior to the mass losing one.

A gas torus can interact with a planet exterior to it as well, so that the angular
momentum extracted from the inner planet can be transferred to the outer planet,
driving it to larger semi-major axes. Thus, we expect the model presented here
to widen the gap between planets that bracket the gaseous torus.

To estimate the approximate scale of this effect, consider two planets of mass
$M_1$ and $M_2$, with semi-major axes $a_1$ and $a_2$
initially. 
If  the inner planet loses a fraction $f_{in}$ of its
mass and  it's eccentricity gets excited to a value $e_{in}$, then the transfer of
the angular momentum would drive the gas torus outwards. However, the presence of a 
second planet can extract angular momentum from the torus, causing it to stall and
the outer planet to move outwards.

In Appendix~\ref{TorusTransfer}, we consider the 
global angular momentum balance between inner planet, torus and outer
planet in this configuration. 
The final semi-major axis ratio is given by
\begin{eqnarray}
&&\left( \frac{a'_2}{a_1} \right)^{1/4}  =  - \frac{f_{in} M_1}{2 M_2} + \left( \frac{a_2}{a_1} \right)^{1/4} \times  \nonumber \\
&&\left[ 1 + \frac{f_{in}^2 (M_1/M_2)^2 + 4 M_1/M_2 (f_{in} + e_1^2/2)}{4(a_2/a_1)^{1/2}} \right]^{1/2} 
\end{eqnarray}
where $a_2/a_1$ is the initial semi-major axis ratio and $a'_2/a_1$ is the final semi-major axis ratio. The torus and
outer planet are assumed to occupy circular orbits.
In the limit of $f_{in}\rightarrow 0$ and $e_{in} \ll 1$, the solution is even simpler, with
\begin{equation}
\frac{P'_2}{P_1} \sim \frac{P_2}{P_1} \left( 1 + \frac{3}{2} e_{in}^2 \frac{M_1}{M_2} \left( \frac{P_2}{P_1} \right)^{-1/3} \right).
\end{equation}
This limit applies when the instantaneous mass in the torus is much smaller than the planetary masses -- it does not
serve as a significant repository of angular momentum, but facilitates the transfer of angular momentum between inner
and outer planet. There is also a natural saturation of this mechanism at $P'_2/P_1 \sim 2.3$, as the strength of the
coupling between planets and gas torus weakens as the gap expands.

\subsection{The Kepler Sample}

To test this model, we have compared the period ratios of Kepler pairs, drawn from the same sample \citep{BHG20}
as in Figure~\ref{Tides}. We focus on the Kepler sample because our comparison will be between planets above
and below the radius gap, and the majority of the planets added from K2 and TESS lie above the gap \citep{BSG23}.
We restrict our comparison to planets with $R< 6 R_{\oplus}$, to avoid the effects of giant planets.

The radius gap defined by the CKLS \citep{FPH17} covered the range 1.5--2 $R_{\oplus}$, which included a sub-population
within the gap. Subsequent analyses \citep{FP18,VAL18,PRI22,HV23} report similar features, included dependences on mass and irradiation. 
 Furthermore,
the radii reported for individual planets can shift by 10\% between different analyses \citep[e.g.,][]{BHG20,BSG23}, so we choose
to calibrate the location of
the radius valley again, to be precise. Our focus is on separating populations that have lost their envelopes relative to those that
have not, so we wish to focus on the location of the minimum, preserving the edges of the gap within their respective
populations.
To do so, we define a narrow window in stellar radius (with width $0.25 R_{\oplus}$) to define the center
of the valley and move the location of this window, to identify the values which provide the most empty gap.
We adopt a slope for the center of the valley that scales with a factor $F = (P/10~\rm days)^{-0.11}$, as has been
determined to best describe the location of the valley \citep{HV23,GPE25}.

For the sample of \cite{BHG20}, the final choice of window  describes a strip that is normalised,
 at $P=10$~days, to the radii  1.81--2.06 $R_{\oplus}$.
We define
our sample of ``Gap-Straddling Pairs'' (GSP) as those for which the inner planet has $R_{in}$ interior to
this strip and
the outer planet has $R_{out}$ exterior to it. To be confident in this division, we consider only planets whose
radii are measured with an absolute $1\sigma$ accuracy $<0.25 R_{\oplus}$ (the width of the gap). 
In order to be consistent with the concept of photoevaporative mass loss, we only include the innermost such pair
in each given system.
This yields
a sample of 136 pairs.

 As a 
comparison sample, we consider neighbouring pairs for which both planets have radii above the valley, with
the same cut in absolute error. This leaves us with 97~pairs.  We call these the ``Outer pairs-in-a-pod'' (OPP), as they should represent the 
properties of the source population prior to any envelope mass loss, and are thought to form a fairly
homogeneous group \citep{WMP18}.    We also define a sample of  ``In-Gap Pairs'' (IGP),
where the inner member of the pair actually lies within the gap,  and the
outer member lies above it. This comprises only 13 pairs, because our  radius window was chosen to minimise this number.
 Finally, we define the sample "Inner pairs-in-a-pod" (IPP), containing neighbouring pairs of planets, both of whom have
radius below the valley. This comprises 212 pairs.

\begin{figure}
\centering
\includegraphics[width=1.0\linewidth]{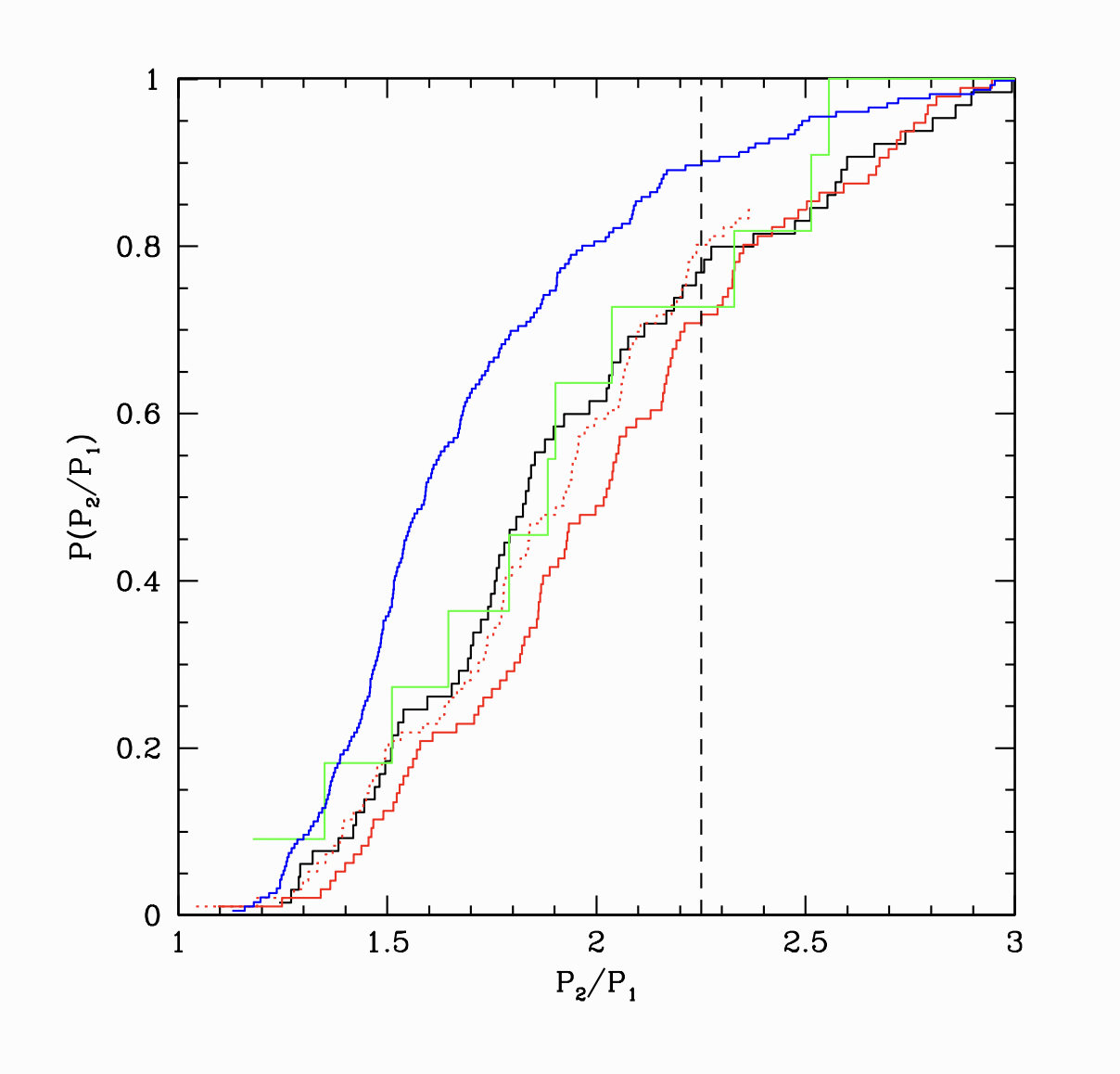}
\caption{The solid red histogram shows the distribution of nearest neighbour period ratios for the
``Gap-Straddling Pairs", in which the inner member has a radius below the radius gap and the
outer member lies above it. The solid black histogram represents the ``Outer pairs-in-a-pod'' sample,
in which both members of the pair lie above the radius valley. In both of these samples, all planets
have absolute radius measurement errors $<0.25 R_{\oplus}$. The dotted red histogram shows
the same as the solid histogram, but with each value divided by a factor 1.05. The solid
green histogram shows the period ratio for pairs in which the inner planet lies within the radius
gap and the outer planet lies outside it.  The blue histogram shows the ``Inner pairs-in-a-pod'' sample. The vertical dashed line shows a value $1.5^2=2.25$.
 \label{PratDis}}
\end{figure}

Figure~\ref{PratDis} shows the cumulative distribution of nearest neighbour period ratios for each
of these samples. The first thing to notice is the distinct mismatch between the green (IPP) and black (OPP)
histograms -- the probability that these two samples are drawn from the same distribution is only $p=2 \times 10^{-6}$.
The fact that these two populations have different spacings suggests 
 that not every planet below the radius gap is simply the former core of a sub-Neptune that
has been stripped of its Hydrogen by photoevaporation or core-powered mass loss. Therefore,  one should not expect 
that every planet below the radius gap has had its eccentricity excited during a mass loss episode.

However, if we assume that the ``peas-in-a-pod'' paradigm \citep{WMP18} is ubiquitous, then we expect that the
planets in any given system should initially have similar compositions. In this picture, the
inner members of the GSP sample should represent the stripped cores of planets that initially had gaseous envelopes.
Indeed, for
 values $P_2/P_1<2.4$, the distribution of GSP period ratios (red histogram) appears to be systematically
shifted to higher values, relative to the OPP distribution (black histogram), and much wider than that of the IPP sample.
The shift, relative to the OPP distribution, is approximately a factor $\sim 1.05$ for $P_2/P_1<1.5$
and closer to 1.1 for $1.5<P_2/P_1<2.1$. 
 Furthermore, the two
distributions converge above $\sim 2.4$, in line with our expectations as to the extent to which
angular momentum can be transferred. Together this result suggests that many of these pairs began as part of
the OPP distribution and have been widened during
the episode of mass loss. A KS-test indicates that these two samples could be drawn from the same
underlying distribution with a probability $p \sim 0.018$ (so, this is a 2.3 $\sigma$ difference).

\begin{figure}
\centering
\includegraphics[width=1.0\linewidth]{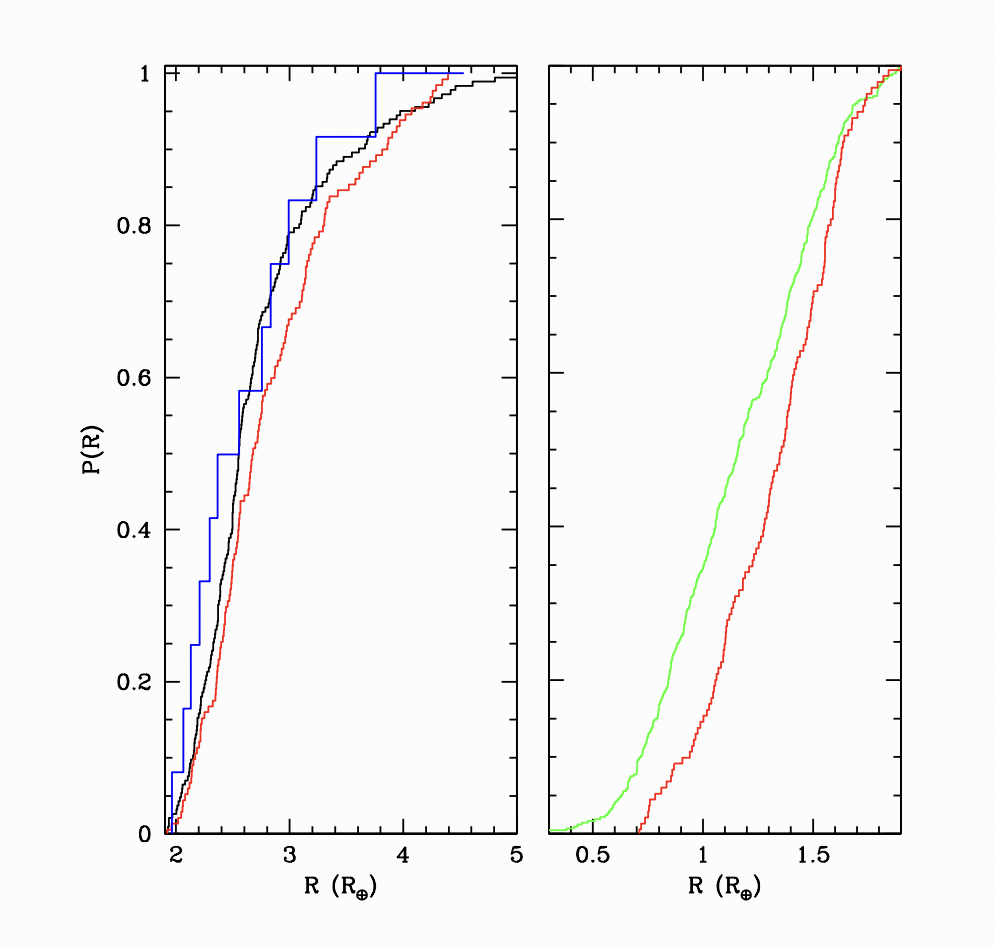}
\caption{The black histogram, in the left hand panel, shows the cumulative distribution of radii for the members of the OPP sample,
while red histogram shows the radius distribution for the outer members of the GSP sample. The blue histogram
shows the same for the IGP sample.  The right hand panel shows the radius distributions below the radius gap,
where the green histogram shows the radius distribution for the IPP sample, compared to the radius distribution of the
inner planets of the GSP sample. This latter sample is biased high compared to the green histogram. The lower limit
of the distributions in the left hand
panel is normalised at $1.9 R_{\oplus}$, while the upper limit of the distributions in the right hand panel is normalised
at the same radius.
 \label{Raddis}}
\end{figure}

The distribution of the planetary radii amongst this sample is also instructive, shown in Figure~\ref{Raddis}. In
the left panel, we show the radii of the planets in the OPP sample (black histogram) -- including both members --
as well as the outer members of the GSP (red) and IGP (blue). The red and black histograms are consistent
with radii drawn from the same parent distribution with a probability $p=0.12$, indicating that the outer planets
of the GSP sample skew slightly to larger radii as well, compared to a generic pair above the radius gap.
The outer planets of the IGP planets are consistent with both, although
too few for any meaningful conclusions.
The
right panel shows the radius comparison of the inner members of the GSP (red) with the distribution 
of radii for the IPP (green histogram) -- including both members.  These distributions are notably inconsistent,
with a probability of only $p = 7\times 10^{-5}$ that they are drawn from the same underlying sample. Thus,
the inner members of the GSP systems are not simply drawn from a parent distribution of IPP planets
-- they are significantly biased to higher values, as would be expected if they are stripped cores of 
planets drawn from the OPP sample.

Other authors have also noted unique properties of planets in or near the gaps in multiplanet systems.
\cite{CB04} show that planet pairs with members in the radius gap have a lower incidence of radius
ratios near unity, which argues that this sub-population contains members whose envelopes have 
been stripped. The sample of \cite{CB04} is not exactly the same as ours, as they include pairs
from K2 and TESS and we require only that our planetary systems straddle the gap, not have a 
member therein.  Nevertheless, both studies support the notion that the radius gap is associated
with an episode of mass loss for the inner planet.
In a comparison of statistical metrics amongst the multi-planet sample, \cite{RSV4} find that systems with a higher gap
complexity \citep{GF20} have fewer sub-Neptunes and more super-Earths.  In principle, this result supports our
results in that mass loss will move planets from the former sample to the latter, and the orbital
evolution described here will increase the gap complexity for those systems that initially began
as equally spaced. However, high values of  gap complexity arise from systems with large variations in
orbital spacing, and we anticipate that the strongest effects will be in systems with compact configurations.

In summary, we note that the observations are consistent with the model in which planet pairs expand when the
inner planet loses mass. For systems where the final planet radius  lies below the radius gap, this shift is consistent
with mass loss fractions of a few percent. For those planets whose final radius lies within the radius gap, the shift is
larger, suggesting mass loss fractions closer to 10\%.  

\section{Conclusions}
\label{Discuss}

We consider the fact that the evaporation of planetary atmospheres can remove the mass from a planet but leave the escaped material
on a heliocentric orbit. Thereafter, gravitational torques between the resulting gaseous torus and the planet can remove
angular momentum from the planet, thereby exciting the planetary eccentricity. We have estimated the plausible range of
eccentricities that can be generated by this process, given that the angular momentum extraction will be limited
 by the mass in the gaseous torus, and the distance out to which
Lindblad resonances can extract angular momentum. We find that this process  is capable of generating eccentricities $e \sim 0.1$, as
has been inferred by \cite{GPE25}. 

We have also considered the effects of this process when it occurs in multiple planet systems. In this case, a gaseous torus
in between two planets can facilitate the exchange of angular momentum between neighbouring planets, resulting in the
widening of the gap. Since the gaseous torus is no longer the ultimate repository of angular momentum in this case, the amount
of extracted angular momentum is potentially larger, although the expansion width is limited by the weakening of the coupling
between torus and planets as the separation expands.

 Data from the Kepler exoplanet sample supports this model, wherein pairs which straddle the radius gap appears to follow a similar 
 period ratio distribution to those
that lie fully above the gap, but shifted outwards by a few percent. The radii of the inner planets in these gap-straddling pairs are
also larger, on average, than the radii of planets in pairs that lie wholly below the gap. This also supports the notion that these
planets are the stripped cores of more massive planets. 

The period ratio distributions of pairs that wholly above and below the radius gap are also notably different, suggesting that
only a fraction of the planets below the radius gap are drawn from stripped remnants of more massive planets. This is
important for the context of this model, because it implies that not every lower radius planet is expected to have eccentricity
excited during an episode of mass loss.  Furthermore, as noted in Figure~\ref{Tides}, many of these planets are not expected
to have tidally circularised, so their lower eccentricities are a remnant of their mechanism of formation. 

Taken together, these observations  argue
 that the elevated eccentricity of planets in or near the radius gap is directly related to the loss of
 the gaseous envelope that drives the evolution of the radius.
  Other models, such as the excitation
of eccentricities as the the result of dynamical instability and planetary collisions \citep{VGH20,GPE25}, predict a
broader range of eccentric planets \citep{CB04}, as the products of collisions should be present
 at a range of planetary radii, not constrained to a narrow range near the radius gap.

It has been suggested that mass loss can itself trigger dynamical instability if it occurs in compact, multiple-planet
systems \citep{WL23}, potentially combining the two above scenarios.
However,  simulations suggest that this requires mass loss fractions of 10--20\%, which is larger
than what is anticipated in most models for the formation of the radius gap. The models of \cite{WL23} also
assumed the mass was accreted onto the host star and little angular momentum was returned to the system.
We have shown that, if the gaseous torus, formed from the lost envelope, interacts with other planets as well,
it can increase period ratios of Gap-spanning planetary pairs by up to 20\%. 

In conclusion, the stripping of a gaseous planetary atmosphere can cause excitation of the eccentricity of the planetary
remnant, if the gas remains in the system long enough to absorb angular momentum through its interaction
with the Lindblad resonances associated with the planetary orbit. This can also drive the expansion of close pairs of
planets, which is supported by the observed properties of Kepler pairs that span the radius gap.

{\bf Data availability}: The data underlying this article will be shared on reasonable request to the corresponding author.

This research has made use of NASA's Astrophysics Data System Bibliographic Services. 
This research has made use of the NASA Exoplanet Archive, which is operated by the California Institute of Technology, under contract with the National Aeronautics and Space Administration under the Exoplanet Exploration Program. This research has made use of NASA's Astrophysics Data System Bibliographic Services.

\bibliographystyle{mnras}
\bibliography{refs}

\appendix

\section{A simple model of angular momentum balance}
\label{TorusTransfer}

We provide here a simple model for the angular momentum transfer between two planets, mediated by interactions with a gaseous torus in between.

If the torus contains a mass that is a fraction $f$ of the final inner planet mass $M_1$, then the original total angular momentum is just that of two neighbouring
planets on circular orbits
\begin{equation}
L_{tot} = (1+f) M_1 \left( G M_* a_1 \right)^{1/2} + M_2 \left( G M_* a_2 \right)^{1/2} \label{Ltot0}
\end{equation}

As gravitational interactions transfer angular momentum from the planetary orbit to that of the gaseous torus, the torus is pushed outwards and the eccentricity
of the inner planet is excited (as noted before, eccentricity excitation occurs faster that semi-major axis decay, so we will assume $a_1$ is constant).  Thus, the
original angular momentum stored in the inner planet is now disbursed between the planet and the torus, at semi-major axis $a_t$
\begin{equation}
L_{in} = M_1 \left( G M_* a_1 (1 - e_1^2)\right)^{1/2} + f M_1 \left( G M_* a_t \right)^{1/2}
\end{equation}
where we have also assumed that the gas torus is on a circular orbit as well.

The transfer of angular momentum is performed by the excitation of outer eccentric Lindblad resonances. The gravitational interactions between the gaseous torus
and the outer planet can also transfer angular momentum, via inner eccentric Lindblad resonances. So, the increase of $a_t$ may be reduced because the torus, can,
in turn, transfer angular momentum to the outer planet orbit, driving the initial semi-major axis $a_2$ to a new value $a_2'$. Thus, conservation of global angular
momentum requires
\begin{equation}
L_{tot} = M_1 \left( G M_* a_1 (1 - e_1^2)\right)^{1/2} + f M_1 \left( G M_* a_t \right)^{1/2} + M_2 \left( G M_* a'_2 \right)^{1/2} \label{Ltot1}
\end{equation}

If we assume that the equilibrium position of the torus is such that the interactions with inner and outer planets are of similar strength, then we expect that
$a_t/a_1 \sim a'_2/a_t$, or $a_t^2 \sim a_1 a'_2$. If we define the variable $x = (a'_2/a_1)^{1/4}$, we can express the equivalence of equations~(\ref{Ltot0}) and
(\ref{Ltot1}) as
\begin{equation}
x^2 + f \frac{M_1}{M_2} x + \frac{M_1}{M_2} \left[ ( 1 - e_1^2)^{1/2}-1 \right] - x_0^2 = 0
\end{equation}
where $x_0 = (a_2/a_1)^{1/4}$. The solution of this quadratic equation yields the final semi-major axis ratio for the planetary pair, as a function of the mass in the
torus and the
eccentricity excited in the inner planet.

The first thing to notice is that this equation has a valid solution even in the limit $f \rightarrow 0$. This is because angular momentum can be transferred between
the two planets even if there is very little gas in the torus at any given time -- it serves only as an intermediary.
In this respect, it behaves in a similar fashion to the proto-comet population
that mediates the outward migration of Uranus and Neptune in the outer Solar system \citep{FI84}.
Thus, in this limit, 
\begin{equation}
x^2 = x_0^2 + \frac{M_1}{M_2} \left[ 1 - ( 1 - e_1^2)^{1/2} \right]
\end{equation}
In the limit where $e_1 \ll 1$, we get a shift in the period ratio of
\begin{equation}
\frac{P'_2}{P_1} =  \frac{P_2}{P_1} \left( 1 + \frac{3}{2} \frac{M_1}{M_2} \frac{e_1^2}{(P_2/P_1)^{1/3}} \right)
\end{equation}

When we include a finite value for $f$, in the same limit of small $f$ and $e_1$, we have
\begin{equation}
\frac{P'_2}{P_1} =  \frac{P_2}{P_1} \left( 1 + \frac{3}{2} \frac{M_1}{M_2} \left[ \frac{e_1^2}{(P_2/P_1)^{1/3}} - 2 f \right] \right)
\end{equation}
A finite mass in the torus does introduce a lower limit on the value of $e_1$, since it requires a finite amount of angular momentum to be stored in
the torus and therefore implies a minimum amount extracted from the inner planet. The resulting minimum is  $e_1^2 > 2 f (P_2/P_1)^{1/3}$.

\begin{figure}
\centering
\includegraphics[width=1.0\linewidth]{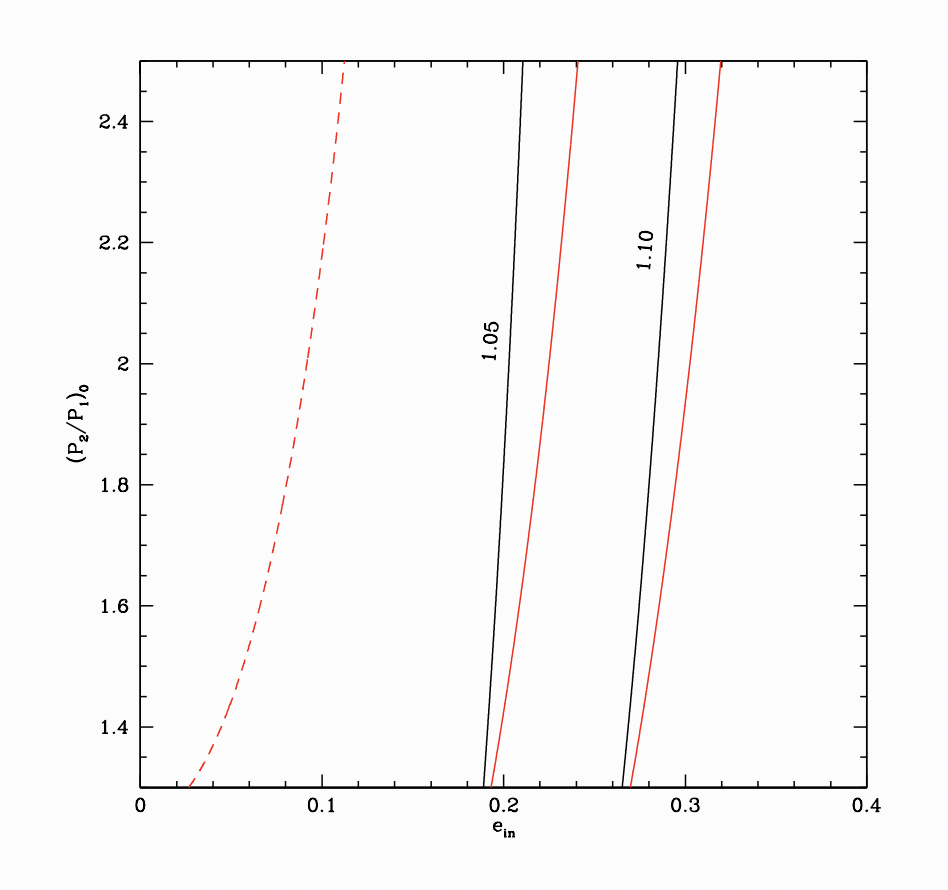}
\caption{The black solid curves show the combination of initial period ratio and final inner eccentricity that
result in a period expansion factor of 1.05 (leftmost) or 1.10 (rightmost), assuming the mass in the torus
is negligible -- assuming an initially equal mass pair. The solid red lines show the equivalent if the mass in the torus is $f_{in}=0.01$ of the innermost
planet. The region to the left of the red dashed line is not accessible in this case while maintaining angular momentum conservation. There
is no such restriction in the case of $f_{in} \rightarrow 0$.
\label{eprat}}
\end{figure}

Figure~\ref{eprat} shows the resulting expansion factors of orbits of different initial $P_2/P_1$
that result from different levels of inner planet eccentricity excitation $e_{in}$. We see that 
an increase in the period ratio of 5--10\% will likely require eccentricity excitation of the inner
planet to levels $e_{in} \sim 0.2$.  Observable levels today  are likely to be approximately half this value,
as secular interactions between the planet pair will excite the eccentricity of the outer planet of the
pair as well, at the expense of the inner. Present day eccentricities may also be damped if the planets
are secularly coupled to additional planets in the system that lie closer to the star and experience
tidal damping \citep{HM15}.

As expressed here, there is potentially no limit to the expansion -- any ratio of $P'_2/P_1$ could be achieved if $e_1$ is large enough.
However, as discussed in  \S~\ref{Torus}, the coupling of the gas torus to the planets operates through the excitation of eccentric Lindblad
resonances, which become weaker at lower order (and greater distance from the planet). This will weaken the coupling between the planets
and serve to limit the degree of orbital expansion. As in \S~\ref{Torus}, if the width of the torus is comparable to the Hill sphere, then we expect
that the coupling weakens for $m<5$, and so we expect that the gas torus will become less effective at driving planets apart once
their period ratio exceeds $\sim (3/2)^2 \sim 2.25$.
Our expectation is that the effects of mass-loss driven expansion
will be more important for compact orbits, and less effective for the wider orbits.

\end{document}